\documentclass[12pt]{article}

\begin{document}

\title{Dephasing by two-level systems at zero temperature by unitary evolution}

\author{Marco Frasca \\
        Via Erasmo Gattamelata, 3, \\
        00176 Roma (Italy)}

\date{\today}

\maketitle

\abstract{We analyze the unitary time evolution of a conduction electron, 
described by a two-level system, interacting with two-level systems (spins)  
through a spin-spin interaction and prove that coherent spin states 
of the conduction electron are obtained
in the strong coupling regime, when the number of the spins
is taken to be, formally, infinitely large (thermodynamic limit).
This model describes a spin interacting with a spin-bath in a strong coupling
regime and gives a dephasing time at zero temperature 
that agrees with the recent experimental results for quantum dots. 
Dephasing is proved to occur as the conduction electron oscillates between the two
states with frequency going to infinity in the thermodynamic limit, that is, increasing the number of
spins. Then, it is shown that the only meaning that can be attached to such oscillations 
with infinite frequency is by an average in time, eliminating the off-diagonal terms of the
density matrix. This model is in agreement with a recent proposal of appearance of classical states  
in quantum mechanics due to the large number of components of a quantum system, 
if properly prepared for the states of each component part.
The strong coupling study of the model is accomplished through the principle of duality as
recently introduced in perturbation theory [M.Frasca, Phys. Rev. A {\bf 58}, 3439 (1998)].
}

\section{Introduction}

Recent experiments have shown that decoherence at zero
temperature can happen in mesoscopic devices as
quantum dots and nanowires \cite{mo,fe}. Indeed, standard
Landau's theory of Fermi liquids disagrees with such findings and rather agrees
with the point of view that, missing any possible exchange of energy between a
system and a reservoir, no decoherence should appear at $T=0$ and
any dephasing time $\tau_\phi$ should go to infinity as 
the temperature goes to zero. Experimentally,
it is observed that such a time saturates at well defined values rather than to go
to infinity as the temperature decreases. Besides, Ferry et al. \cite{fe} have
shown that, for quantum dots, the dephasing time saturates going like the inverse
of the number of the electrons in the dots. 


Several attempts to explain such a disagreement between theory and experiment
have been put forward. An approach discussed in Ref.\cite{gz} explains
the dephasing as an effect of zero-point fluctuations being these the only residual of
a quantum system at zero temperature but this approach has undergone severe criticism
in Ref.\cite{cri}. Vacuum fluctuation effects on dephasing was also studied in Ref.\cite{butt}. 
In Ref.\cite{alt1,2ck} has been proposed to
consider the effect of two-level tunneling systems (TLS), representing the impurities in
the system, as the cause of dephasing: Recent computation by Mohanty et al. and
Altshuler et al.\cite{alt2} have shown that the density of impurities needed to
have the right order of magnitude for the dephasing time is larger than what
is found in mesoscopic devices. In the same line of considering TLS as the
origin of dephasing, it has been proposed that a two-channel Kondo interaction\cite{2ck} 
can be the reason of the saturation. A critical point in this proposal is the value of the
Kondo temperature. An evaluation of the Kondo temperature in mesoscopic systems
has been given in \cite{alt3}.

The aim of this paper is to give an analysis of a model
in the strong coupling regime for TLSs coupled to a conduction electron, and
prove that the dephasing is just a property of the unitary evolution of the
model in the ``thermodynamic limit'' where the number of the TLS is formally taken to
be increasingly large. This is in agreement
with a recent proposal for decoherence as originating from unitary evolution
of N quantum systems with N becoming increasingly large \cite{fra}. The main point of
this approach is that it depends on the way the system is prepared. When quantum
fluctuations can be neglected with respect to the mean values 
then, by the Ehrenfest theorem the system
follows the classical equations of motion without no significant deviation.

The model we analyze has been also discussed, in a different regime and with
some minor differences by H\"anggi and Shao \cite{hs}, but here we apply the
strong coupling analysis and a different decoherence mechanism to achieve
agreement with experiments on quantum dots. Our decoherence mechanism is non
dissipative, being derived by the unitary evolution, taking the thermodynamic
limit. A similar kind of effect has been recently uncovered in a
generalization of the model we consider here \cite{loss},
but considering a coupling for all the spin components. Decoherence
is produced dynamically and no arbitrary division between bath and system
has to be done.

We will show that, modelling the conduction electron by a two-level system, with 
a spin-spin term in the strong coupling regime for the interaction with TLSs, 
the above scheme for decoherence applies 
and the correlators for the conduction electron go to zero in the thermodynamic limit,
that is, increasing the number of spins without bound. It is
interesting to point out that we are considering the case of strong coupling
where small perturbation theory does not apply contrarily at the analysis of
Mohanty et al. \cite{alt2}. What we will prove is that the conduction electron
evolves in time as a coherent spin state (CSS)\cite{css,gil}, that is, a state 
having mean values following the classical equations of motion and the
uncertainty product at the minimum. 
These states are SU(2) analogous to the coherent states describing a laser
field that derives from bosonic operators. What we will show is that the conduction
electron, being in the ground state, evolves in time with a coherent spin state
when the number of interacting spins is taken to become very large\cite{gil}. The electron
oscillates between the two states it can access with frequency going to infinity in this
limit (thermodynamic limit). A system having a time-scale going to zero or, that
is the same, an infinite frequency of oscillations can have physical meaning
only if it is averaged in time as we will prove.

The dephasing time we obtain goes like the inverse of the number of spins in the
bath and this agrees with the experimental results given in \cite{fe} where the
spins involved are those of the 2D electron gas in the dot.

In order to study this model in the strong coupling regime we apply the dual
Dyson series \cite{fra1}. This approach has been pioneered by
Bender et al. \cite{ben} in quantum field theory. These results can be framed 
in a general formulation given
in Ref.\cite{fra1} that permits to prove the main assert of this paper. 

The paper is so structured. In Sec.\ref{sec1} we introduce the model of
spin-spin interaction. In Sec.\ref{sec2} we give a description of the method
of strong coupling perturbation theory or dual Dyson series. In Sec.\ref{sec3}
the perturbative solution of the model in the strong coupling regime is
given. In Sec.\ref{sec4} we give the conclusions with some discussion of the
theoretical results we obtain. 
Finally, in two appendices we discuss about the model of non dissipative
decoherence we use and the coherent spin states.

\section{A model for the conduction electron in a mesoscopic device}
\label{sec1}

With small modifications with respect to the model of Ref.\cite{hs} we take
for the conduction electron (here and in the following $\hbar=1$)
\begin{equation}
    H_{el}=\frac{\Omega_0}{2}\sigma_z
\end{equation}
being $\Omega_0$ the energy separation 
of the ground and the excited states. A two-level approximation
proves to be a good one also for the study of Kondo models \cite{alt3,cox}. Besides,
devices for representing qubit have also been devised in Ref.\cite{seth}. To fix
the ideas we assume that we are treating the electron spin but, any other property
that can be characterized by two eigenvalues can be used for the following analysis.

In the same way one can write the Hamiltonian of TLSs representing the spin bath
\begin{equation}
    H_{TLS}=\frac{1}{2}\sum_{i=1}^N (\Delta_{xi}\tau_{xi}+\Delta_{zi}\tau_{zi}),
\end{equation}
being $\tau_{xi}$,$\tau_{zi}$ the Pauli matrices for the $i$-th TLS, $\Delta_{xi}$ 
and $\Delta_{zi}$ the parameters of the spins belonging to the bath.

Finally, we hypothesize that the TLSs and a conduction electron interact through
an spin-spin term, that is \cite{hs}
\begin{equation}
     H_K=-J\sigma_x\cdot\sum_{i=1}^N \tau_{xi}.
\end{equation}
being $J$ the strength of the coupling. 
This kind of approximation is used e.g. in Josephson-junction devices \cite{seth,jjd}
where tunneling is the main effect. Dephasing by TLSs has been suggested also in this
case \cite{nar}. So, finally, we write the Hamiltonian that we want to analyze as
\begin{equation}
    H=H_{el}+H_{TLS}+H_K=\frac{\Omega_0}{2}\sigma_z+
	\frac{1}{2}\sum_{i=1}^N (\Delta_x\tau_{xi}+\Delta_z\tau_{zi})
	-J\sigma_x\cdot\sum_{i=1}^N \tau_{xi}. \label{eq:model}
\end{equation}
We will assume that the interaction and the electronic terms will prevail
on the TLS Hamiltonian. Besides, the initial state of TLSs can be cast in the form
\begin{equation}
    |\chi_0\rangle=\prod_{i=1}^N|\lambda_i\rangle
\end{equation}
with $\tau_{xi}|\lambda_i\rangle=\lambda_i|\lambda_i\rangle$, being 
$\lambda_i=\pm 1$ and also
\begin{equation}
    |\lambda_i\rangle=\frac{1}{\sqrt{2}}\left[
	\left(\begin{array}{c} 0 \\ 1 \end{array} \right)_i+
	\lambda_i
	\left(\begin{array}{c} 1 \\ 0 \end{array} \right)_i
	\right].
\end{equation}
Our aim will be the study of this model in the strong coupling regime, that is,
when the interaction term is larger than the TLS Hamiltonian as is the term of
the conduction electron. This is the crucial approximation we take in this paper.

\section{Duality in perturbation theory}
\label{sec2}

The concept of duality in perturbation theory has been introduced in Ref.\cite{fra1}.
This idea embodies the pioneering work done by Bender at al.\cite{ben} in
quantum field theory where the kinetic term is taken as a perturbation and use
is made of path integral formalism.

This approach can be easily formalized by considering the Schr\"odinger equation
\begin{equation}
    (H_0+\lambda V)|\psi(t)\rangle=i\frac{\partial}{\partial t}|\psi(t)\rangle
	\label{eq:sch}
\end{equation}
being $\lambda$ a parameter, and $H_0$ and $V$ indicate two parts by which the
original Hamiltonian has been split. Our aim is to use the freedom in the
choice of the two parts to obtain two different perturbation series. Indeed, if
$\lambda\rightarrow 0$ one obtains the Dyson series
\begin{equation}
    |\psi(t)\rangle=e^{-iH_0t}Te^{-i\lambda\int_0^tdt'V_I(t')}|\psi(0)\rangle \label{eq:ds}
\end{equation} 
being $T$ the time ordering operator and
\begin{equation}
    V_I(t)=e^{iH_0t}Ve^{-iH_0t},
\end{equation}
having assumed both $H_0$ and $V$ time independent.

For $\lambda\rightarrow\infty$, we rescale time as $\tau=\lambda t$ so that
eq.(\ref{eq:sch}) becomes
\begin{equation}
    \left(\frac{1}{\lambda}H_0+V\right)|\psi(\tau)\rangle=i\frac{\partial}{\partial\tau}|\psi(\tau)\rangle
\end{equation}
and so, by this rescaling, we have reverted the role of $H_0$ and $V$ giving a
Dyson series with the development parameter $\frac{1}{\lambda}$
\begin{equation}
    |\psi(t)\rangle=e^{-iV\tau}Te^{-i\frac{1}{\lambda}\int_0^{\tau}d\tau'H_{0F}(\tau')}|\psi(0)\rangle
\end{equation}
being now
\begin{equation}
    H_{0F}(\tau)=e^{iV\tau}H_0e^{-iV\tau}.
\end{equation} 
This perturbation series is dual to the series eq.(\ref{eq:ds}) being its development
parameter $\frac{1}{\lambda}$ that is the inverse of the one of the series eq.(\ref{eq:ds}).
But this has been obtained by a symmetry of the original Hamiltonian where one part or
the other can be chosen arbitrarily: This is duality in perturbation theory. Indeed,
a series can be obtained from the other by the interchange $H_0 \leftrightarrow V$,
setting $\lambda=1$. It is interesting to note that, if the initial condition is e.g.
an eigenstate of $H_0$, the Dyson series gives a trivial leading order yielding the
same eigenstate multiplied by a phase factor while the dual Dyson series gives a
non trivial one being the exponential factor $e^{-iV\tau}$ multiplied 
by the eigenstate of $H_0$. The argument can be reversed but this simmetry 
in the perturbation series can be
broken by the way a quantum system is 
initially prepared. In quantum mechanics, also for
strong perturbations, it is always taken as initial state the one of the
unperturbed system. 

It is quite easy to recover the result of Bender et al.\cite{ben} by taking for
$V$ the kinetic term of a $\lambda\phi^4$ theory.

It is interesting to point out here that some condition on the boundedness of the
domains of the operators $H_0$ and $V$ should be extended from the Dyson series
to its dual as some of the most important operators in quantum mechanics are
unbounded. In the following this problem will be of no concern as our operators
act on a finite dimensional Hilbert space.

\section{Solution in the strong coupling regime}
\label{sec3}

Our aim is to show how non dissipative decoherence, as described in
Appendix \ref{app1}, appears by the dynamics of model (\ref{eq:model}). We will
see that coherent spin states are the solution at the leading order (see Appendix
\ref{app2} for a description of spin coherent states).

From the Hamiltonian (\ref{eq:model}) we can easily get the strong coupling
expansion by taking
\begin{eqnarray}
    H_0&=&\frac{1}{2}\sum_{i=1}^N (\Delta_{xi}\tau_{xi}+\Delta_{zi}\tau_{zi}) \\
	V &=& \frac{\Omega_0}{2}\sigma_z-J\sigma_x\cdot\sum_{i=1}^N \tau_{xi}
\end{eqnarray}
and the dual Dyson series is given by
\begin{equation}
    U_D=U_0(t)\left[1-i\int_0^tH_{0F}(t')dt'-\int_0^tdt'\int_0^{t'}dt''H_{0F}(t')H_{0F}(t'')+\cdots\right]
\end{equation}
where
\begin{equation}
    U_0(t)=\exp\left[-it\left(
	       \frac{\Omega_0}{2}\sigma_z-J\sigma_x\cdot\sum_{i=1}^N \tau_{xi}
		   \right)\right]
\end{equation}
and
\begin{equation}
    H_{0F}(t)=U_0^\dagger(t)
	          \left[\frac{1}{2}\sum_{i=1}^N (\Delta_{xi}\tau_{xi}+\Delta_{zi}\tau_{zi})\right]
			  U_0(t)
\end{equation}
that is, a really non trivial result already at the leading order if the initial state
is an eigenstate of $H_0$ giving in this case $U_0(t)$ multiplied by this same state.

Firstly, let us evaluate the leading order result. By the disentanglement formula
(\ref{eq:dis}) of Appendix \ref{app2}, we obtain immediately
\begin{equation}
    U_0(t)=\exp\left(-i\hat\Lambda(t)\sigma_+\right)
	       \exp\left(-\ln\hat\Sigma(t)\sigma_3\right)
		   \exp\left(-i\hat\Lambda(t)\sigma_-\right)
\end{equation}
being the operators
\begin{eqnarray}
    \hat\Lambda(t)&=&\frac{-J\sum_{i=1}^N\tau_{xi}}{\hat\Omega}
	                 \frac{\sin(\hat\Omega t)}{\hat\Sigma(t)} \\
    \hat\Sigma(t)&=&\cos(\hat\Omega t)+i\frac{\Omega_0}{2\hat\Omega}\sin(\hat\Omega t) \\
	\hat\Omega(t)&=&\sqrt{\frac{\Omega_0^2}{4}+J^2\left(\sum_{i=1}^N\tau_{xi}\right)^2}.
\end{eqnarray}
We note that these operators are diagonal on the environment state and so we have to compute\cite{imry}
\begin{equation}
    \bar U_D(t)=\langle\chi_0|U_D(t)|\chi_0\rangle
\end{equation}
that is, assuming e.g. $\sum_{i=1}^N\tau_{xi}|\chi_0\rangle=-N|\chi_0\rangle$, at the leading order
\begin{equation}
    \bar U_0(t)=\exp\left(-i\Lambda(t)\sigma_+\right)
	       \exp\left(-\ln\Sigma(t)\sigma_3\right)
		   \exp\left(-i\Lambda(t)\sigma_-\right)
\end{equation}
and
\begin{eqnarray}
    \Lambda(t)&=&\frac{NJ}{\Omega}\frac{\sin(\Omega t)}{\Sigma(t)} \\
    \Sigma(t)&=&\cos(\Omega t)+i\frac{\Omega_0}{2\Omega}\sin(\Omega t) \\
	\Omega(t)&=&\sqrt{\frac{\Omega_0^2}{4}+(NJ)^2}.
\end{eqnarray}
Instead, the first order term is given by
\begin{equation}
    -i\bar U_0(t)\int_0^t \langle\chi_0|H_0|\chi_0\rangle dt'
\end{equation}
that is,
\begin{equation}
    i\bar U_0(t)\frac{N}{2}\bar\Delta_x t
\end{equation}
a secular term with $\bar\Delta_x=\frac{1}{N}\sum_{i=1}^N\Delta_{xi}$, 
unbounded in the limit $t \rightarrow\infty$. This terms can be
resummed (see Ref.\cite{fra2}) to give an exponential correction to the leading
order $\exp\left(i\frac{N}{2}\bar\Delta_x t\right)$.

At the second order one has
\begin{equation}
    -\bar U_0(t)\int_0^tdt'\int_0^{t'}dt''\bar U_0^\dagger(t')
	\langle\chi_0|H_0U_0(t')U_0^\dagger(t'')H_0|\chi_0\rangle\bar U_0(t'').
\end{equation}
To compute this term we note that
\begin{equation}
    \sum_{i=1}^N\Delta_{zi}\tau_{zi}|\chi_0\rangle=\sum_{i=1}^N\Delta_{zi}|\chi_{0i}\rangle \label{eq:state}
\end{equation}
where
\begin{equation}
    |\chi_{0i}\rangle=|-1\rangle_1 |-1\rangle_2\cdots |1\rangle_i\cdots |-1\rangle_N
\end{equation}
that is, (\ref{eq:state}) is a non normalized state orthogonal to $|\chi_0\rangle$. Then, after some
algebra one gets
\begin{eqnarray}
    -\frac{1}{2}\frac{N^2}{4}\bar\Delta_x^2 t^2\bar U_0(t)
	&-&\frac{1}{4}\sum_{i=1}^N\Delta_{zi}^2\bar U_0(t)
	\int_0^tdt'\int_0^{t'}dt''e^{it'\left(\frac{\Omega_0}{2}\sigma_z+NJ\sigma_x\right)}
	                          e^{-it'\left[\frac{\Omega_0}{2}\sigma_z+(N-1)J\sigma_x\right]}\times \\
	 & &                      e^{it''\left[\frac{\Omega_0}{2}\sigma_z+(N-1)J\sigma_x\right]}
	                          e^{-it''\left(\frac{\Omega_0}{2}\sigma_z+NJ\sigma_x\right)}.
\end{eqnarray}
The first term is just the second order term of the Taylor series
that enters into the resummation of $\exp\left(i\frac{N}{2}\bar\Delta_x t\right)$ 
and can be eliminated.

If $N\gg 1$ the second term can be evaluated. It gives
\begin{equation}
    -\frac{1}{4}\frac{\sum_{i=1}^N\Delta_{zi}^2}{J^2}\bar U_0(t)
	\left(1-e^{iJt\sigma_x}+iJt\sigma_x\right)
\end{equation}
and again we obtain a secularity. This can be resummed as done for the other one
but we do not pursue this aim here. Rather, we note that for consistency reasons
we need to have $\frac{\sum_{i=1}^N\Delta_{zi}^2}{J^2}\ll 1$, otherwise we miss
convergence. This condition should be kept in the thermodynamic limit 
$N\rightarrow\infty$ and realizes the feature of this strong coupling
expansion at this order. It is interesting to note that the contribution $\Omega_0$
of the conduction electron plays no role in the thermodynamic limit
and this is in agreement with the experimental results given in \cite{fe}. 
Indeed, in such a case we have
\begin{equation}
    \bar U_0(t)\approx\exp\left(-itNJ\sigma_x\right)
\end{equation}
that using the relation $\sigma_x=\sigma_++\sigma_-$ proves to be the operator
generating a coherent spin state as given in eq.(\ref{eq:cs}) being now
$\zeta=-itNJ$. If the conduction electron is in the extremal state 
$\left(\begin{array}{c} 0 \\ 1 \end{array} \right)$, we obtain Rabi oscillations
with frequency $\Omega_R=2NJ$.

Our aim is to see if, in the limit $N\rightarrow\infty$, a meaning can be attached
to the above evolution operator. We are requested to give a sense to the limits
\begin{eqnarray}
    \lim_{N\rightarrow\infty}\cos(Nx) \\
	\lim_{N\rightarrow\infty}\sin(Nx)
\end{eqnarray}
and this can be done if we rewrite the above as the values of divergent series.
Indeed, one has
\begin{eqnarray}
    \cos(Nx)&=&1-\int_0^{Nx}\sin(y)dy \\
	\sin(Nx)&=&\int_0^{Nx}\cos(y)dy
\end{eqnarray}
that we can reinterpret for our aims through the Abel summation to divergent integrals 
as\cite{dist}
\begin{eqnarray}
    \lim_{\epsilon\rightarrow 0^+}\lim_{N\rightarrow\infty}\int_0^{Nx}e^{-\epsilon y}\cos(y)dy \\
	\lim_{\epsilon\rightarrow 0^+}\lim_{N\rightarrow\infty}\int_0^{Nx}e^{-\epsilon y}\sin(y)dy,
\end{eqnarray}
where the order with which the limits are taken is important. These are proper constructions for
the thermodynamic limit in this case and
give $0$ and $1$ respectively so that, the evolution operator in the thermodynamic
limit can be taken to be zero. This is the correct meaning to be attached physically to
functions having time scales of variation going to $0$ (instantaneous variation). 
Indeed, whatever physical
apparatus one uses to make measurements on such a system, unavoidably
an average in time will be made that, in this case, gives zero. 

This conclusion is fundamental for the density matrix that, in this way, turns out
to have the off-diagonal terms averaged to zero and the diagonal ones to $\frac{1}{2}$.
This is decoherence, as promised, in the thermodynamic limit. For the sake of
completeness, we report here the case of the density matrix. We have for the
conduction electron in the ground state
\begin{equation}
    \rho(t)=\exp\left(-itNJ\sigma_x\right)
	\left(
	\begin{array}{clcr}
	0 & 0 \\
	0 & 1
	\end{array}
	\right)\exp\left(itNJ\sigma_x\right)
\end{equation}
that yields
\begin{eqnarray}
    \rho_{1,1}(t)&=&\frac{1-\cos(2NJt)}{2} \\
	\rho_{1,-1}(t) &=& -i\frac{1}{2}\sin{2NJt} \\
	\rho_{-1,1}(t) &=& i\frac{1}{2}\sin{2NJt} \\
	\rho_{-1,-1}(t)&=&\frac{1+\cos(2NJt)}{2}.
\end{eqnarray}
Applying the above argument in the limit $N\rightarrow\infty$ gives the required result.
From the above matrix elements we find a natural time scale for the model for $N\gg 1$,
that is $\tau_0=\frac{\pi}{NJ}$, that, as already said, goes to zero in the
thermodynamic limit.

The differences with the case of quantum dissipative systems \cite{weiss} can be
straightforwardly understood. A quantum dissipative system is generally characterized
by substituting sums with integrals and doing hypothesis on the spectral function for
the energy of the bath. In this way dissipation is recovered in a form of a Langevin
equation. In our case, decoherence appears without dissipation as a dynamical
process by the unitary evolution, solving directly the Schr\"odinger equation for
the many-body problem and checking directly the solution. In this way, no ad hoc
hypothesis is needed on the bath that now is taken together with the system as
a single system whose evolution has to be studied. Decoherence at zero temperature
is then obtained. 

An analog situation for the bosonic case is discussed in Ref.\cite{fra}.  

\section{Discussion and Conclusions}
\label{sec4}

The above analysis shows that, in a strong coupling regime, if the number of
spins belonging to the bath 
becomes increasingly large, one has firstly that the motion of the
conduction electron enters into a coherent spin state oscillating between the
two states we have supposed it can access. Secondly, if the thermodynamic
limit, as defined through the resummation technique of Abel, is taken, then,
the conduction electron is localized in one of the two states with probability
$\frac{1}{2}$. So, the TLSs act like a classical measurement apparatus
erasing the interference terms of the density matrix that, in this way, takes
a mixed form.

Such an approach can give an explanation to dephasing at zero
temperature recently observed in quantum dots\cite{fe}. Indeed, we have
derived a time scale for the oscillations of the conduction electron, that is,
the period $\tau_0=\frac{\pi\hbar}{NJ}$ (having restated the Planck constant).
But this time should be the same of the saturated values experimentally observed,
i.e. a few nanoseconds. This, in turn, means that $NJ\approx k_B T_0$ being
$T_0\approx 0.1 K$ the saturation temperature and so,
the order of magnitude of energy is a few of $\mu$eV. This result is
in agreement with the saturation time for quantum dots with $N$ being the number
of spins (electrons) found in the dot.

The most important conclusion is that in this paper we have given a consistent
set of methods to approach the study of a model in the
strong coupling regime where, in the thermodynamic limit, effects of decoherence
may appear in the zero temperature case. The model agrees with recent experiment
on quantum dots.

\appendix

\section*{Appendices}

\section{Classical states by unitary evolution}
\label{app1}

The main result we present here, based on the results of Ref.\cite{fra}, is that,
$N$ TLSs, in the thermodynamic limit, understood as the formal limit $N\rightarrow\infty$,
behaves, if their states are properly disordered, as a classical system 
and so, a two-level system interacting with such a bath undergoes
decoherence\cite{zur}. The critical point is that the classical behavior of such
N-TLSs system emerges or not depending on the way the system is prepared. The
question pointed out in Ref.\cite{fra,zur} is that the most realistic situation
is the one having the N-TLSs in the state
\begin{equation}
    |\chi_0\rangle=\prod_{i=1}^N(\alpha_i|-1\rangle_i+\beta_i|1\rangle_i) \label{eq:chi}
\end{equation}
being
\begin{equation}
    |\alpha_i|^2+|\beta_i|^2=1
\end{equation}
and
\begin{equation}
    \tau_{zi}|\pm 1\rangle_i=\pm|\pm 1\rangle_i
\end{equation}
being $\tau_{zi}$ the Pauli matrix for the i-th component of the N-TLSs. The TLS Hamiltonian
by itself, without interactions, can be diagonalized and, for the sake of semplicity, we
assume all the parameters to be not dependent on the number $i$ of the given TLS, as, in this way we 
cannot expect the result to change too much. So,
the Hamiltonian of the TLSs can be cast in the form
\begin{equation}
    H=\lambda\sum_{i=1}^N\tau_{zi}.
\end{equation}
having $\lambda = \sqrt{\Delta_x^2+\Delta_z^2}$. Then, the system evolves in time as
\begin{equation}
    |\chi_0(t)\rangle=\prod_{i=1}^N(\alpha_ie^{i\lambda t}|-1\rangle_i+\beta_ie^{-i\lambda t}|1\rangle_i)
\end{equation}
It can be proven that\cite{fra}
\begin{equation}
    \frac{\Delta H}{\langle H\rangle}\propto \frac{1}{\sqrt{N}}
\end{equation}
being $(\Delta H)^2$ the variance computed on the state $|\chi_0(t)\rangle$,
or, that is the same, on the state $|\chi_0(0)\rangle$ and
$\langle H\rangle$ the mean computed on the same state. So,
quantum fluctuations are negligible in the thermodynamic limit. The same
can be said for the spin components $\Sigma_x=\sum_{i=1}^N\tau_{xi}$ and
$\Sigma_y=\sum_{i=1}^N\tau_{yi}$ that obeys the classical equations of motion by the
Ehrenfest theorem
\begin{eqnarray}
    \langle\dot\Sigma_x\rangle&=&-2\lambda\langle\Sigma_y\rangle \\
	\langle\dot\Sigma_y\rangle&=&2\lambda\langle\Sigma_x\rangle
\end{eqnarray}
without any significant deviation due to quantum fluctuations. This is exactly the
behavior of a coherent spin state\cite{gil}: quantum and classical dynamics
coincide when the Hamiltonian is a linear combination of the generators of the
symmetry group from which the coherent states originate and the system is found
in the extremal state, to be defined later, or in a coherent state. In both these 
cases, the evolution of the quantum system is described by a coherent state. It is
important to emphasize that we have described an unitary evolution and no
dissipative effect is really involved. Anyhow, it should be emphasized that taking
the thermodynamic limit is a fundamental step in our approach.

\section{Coherent spin states}
\label{app2}

We limit our analysis to the case of a SU(2) group for a spin $\frac{1}{2}$ 
particle whose generators can be set through the three Pauli matrices 
$\sigma_+$, $\sigma_-$ and $\sigma_z$ having the algebra
\begin{eqnarray}
    \left[\sigma_+,\sigma_-\right] &=& \sigma_z\\
	\left[\sigma_z,\sigma_\pm\right] &=& \pm 2\sigma_\pm.
\end{eqnarray}
Then, a coherent spin state can be defined as \cite{css,gil}
\begin{equation}
    |\frac{1}{2},\zeta\rangle=e^{\zeta\sigma_+-\zeta^*\sigma_-}|\frac{1}{2},-\frac{1}{2}\rangle
	\label{eq:cs}
\end{equation}
being $\sigma_z|\frac{1}{2},-\frac{1}{2}\rangle=-|\frac{1}{2},-\frac{1}{2}\rangle$ 
the lower eigenstate said an extremal or ground state, and $\zeta$ a complex number.
Analogously, one has $\sigma_z|\frac{1}{2},\frac{1}{2}\rangle=|\frac{1}{2},\frac{1}{2}\rangle$. 
If the Hamiltonian is built in such a way to be a linear combination of
the generators of the SU(2) group,
the time evolution of a coherent state gives again a coherent state. If the
initial state is an extremal one, then the state evolves into a coherent state. 
In both these cases, there is no difference between classical and quantum dynamics\cite{gil}.
What really we have done is just to rotate the extremal state on the plane
x-y by an angle $\theta$ around the axis $(\sin\phi,-\cos\phi,0)$. 
This can be realized by setting $\zeta=\frac{\theta}{2}e^{-i\phi}$ into eq.(\ref{eq:cs}).
In terms of the eigenstates $|\frac{1}{2},\frac{1}{2}\rangle$ and
$|\frac{1}{2},-\frac{1}{2}\rangle$ one can write as for bosonic coherent states
\begin{equation}
    |\frac{1}{2},\zeta\rangle=
	\frac{1}{(1+\tau\tau^*)^\frac{1}{2}}\exp{(\tau\sigma_+)}|\frac{1}{2},-\frac{1}{2}\rangle=
	\frac{1}{(1+\tau\tau^*)^\frac{1}{2}}
	\left(|\frac{1}{2},-\frac{1}{2}\rangle +
	      \tau|\frac{1}{2},\frac{1}{2}\rangle
	\right)
\end{equation}
being
\begin{equation}
    \tau=\tan\frac{\theta}{2}e^{-i\phi}=\frac{\zeta\sin{|\zeta|}}{|\zeta|\cos{|\zeta|}}.
\end{equation}
This description of a spin $\frac{1}{2}$ through a rotated state is
a particular case of a more general concept that coherent spin states, to be similar to
a bosonic coherent state of N particles, describe an assembly of N spin $\frac{1}{2}$
particles.

A simple way to generate a coherent spin state is given by a spin $\frac{1}{2}$ 
particle in a constant magnetic field with Hamiltonian given by
\begin{equation}
    H=-\frac{1}{2}g \mu_BB_x\sigma_x-\frac{1}{2}g \mu_BB_y\sigma_y \label{eq:H}
\end{equation} 
that has a time evolution operator
\begin{equation}
    U=\cos\left(\Omega t\right)+i\frac{B_x \sigma_x+B_y \sigma_y}{\sqrt{B_x^2+B_y^2}}
	  \sin\left(\Omega t\right)
\end{equation}
being $\Omega=\frac{1}{2}g \mu_B\sqrt{B_x^2+B_y^2}$. Then, if we apply the
above operator on the extremal state $|\frac{1}{2},-\frac{1}{2}\rangle$ we get
\begin{equation}
    |\frac{1}{2},\zeta\rangle=\cos\left(\Omega t\right)|\frac{1}{2},-\frac{1}{2}\rangle+
	i\frac{B_x-iB_y}{\sqrt{B_x^2+B_y^2}}\sin\left(\Omega t\right)|\frac{1}{2},\frac{1}{2}\rangle.
\end{equation}
So, choosing
\begin{equation}
    \tau=i\frac{B_x-iB_y}{\sqrt{B_x^2+B_y^2}}\tan\left(\Omega t\right)
\end{equation}
we realize that we have built a coherent spin state. It is important to note
that in this case the extremal state is not an eigenstate of the Hamiltonian 
(\ref{eq:H}). Besides, these are minimum uncertainty states for the rotated
components by an angle $\phi$. That is, if we define
\begin{eqnarray}
    \langle\sigma_\eta\rangle&=&\langle\sigma_x\rangle\cos\phi+\langle\sigma_y\rangle\sin\phi \\
    \langle\sigma_\xi\rangle&=&-\langle\sigma_x\rangle\sin\phi+\langle\sigma_y\rangle\cos\phi
\end{eqnarray}
and being
\begin{eqnarray}
    \langle\sigma_x\rangle&=&\cos\phi\sin(2\theta) \\
	\langle\sigma_y\rangle&=&\sin\phi\sin(2\theta) \\
	\langle\sigma_z\rangle&=&-\cos(2\theta)
\end{eqnarray}
then
\begin{eqnarray}
    \langle\sigma_\eta\rangle&=&\sin(2\theta) \\
    \langle\sigma_\xi\rangle&=&0
\end{eqnarray}
and finally
\begin{eqnarray}
    (\Delta\sigma_\eta)^2&=&\cos^2(2\theta) \\
	(\Delta\sigma_\xi)^2&=&1 \\
	(\Delta\sigma_z)^2&=&\sin^2(2\theta).
\end{eqnarray}
This means that the uncertainty relations become
\begin{eqnarray}
    \Delta\sigma_\eta\Delta\sigma_\xi&=&\frac{1}{2}|\langle[\sigma_\eta,\sigma_\xi]\rangle|=|\langle\sigma_z\rangle| \\
	\Delta\sigma_\eta\Delta\sigma_z&=&\frac{1}{2}|\langle[\sigma_\eta,\sigma_z]\rangle|\ge|\langle\sigma_\xi\rangle|=0 \\
	\Delta\sigma_\xi\Delta\sigma_z&=&\frac{1}{2}|\langle[\sigma_\xi,\sigma_z]\rangle|=|\langle\sigma_\eta\rangle|.
\end{eqnarray}
For the special case of $\sigma_\eta$ and $\sigma_z$ these can be seen as classical commuting variables.
To complete this section we give some disentanglement formulas. The following relation is true\cite{bar}
\begin{equation}
     \exp\left(\lambda_+\sigma_++\lambda_-\sigma_-+\frac{\lambda_3}{2}\sigma_3\right)=
	 \exp(\Lambda_+\sigma_+)\exp\left[-\frac{1}{2}(\ln\Lambda_3)\sigma_3\right]\exp(\Lambda_-\sigma_-)
	 \label{eq:dis}
\end{equation}
where
\begin{eqnarray}
    \Lambda_3&=&\left(\cosh\alpha-\frac{\lambda_3}{2\alpha}\sinh\alpha\right)^{-2} \\
	\Lambda_\pm&=&\frac{2\lambda_\pm\sinh\alpha}{2\alpha\cosh\alpha-\lambda_3\sinh\alpha} \\
	\alpha^2&=&\frac{1}{4}\lambda_3^2+\lambda_+\lambda_-.
\end{eqnarray}
For the particular case of $\lambda_\pm=i\theta$ and $\lambda_3=0$ one gets
\begin{equation}
    \exp[i\theta(\sigma_++\sigma_-)]=\exp[i(\tan\theta)\sigma_+]
	                                 \exp\left[-\frac{1}{2}\ln(\cos^2\theta)\sigma_3\right]
									 \exp[i(\tan\theta)\sigma_-].
\end{equation}

\end{document}